\documentclass[amsmath, amssymb, 
superscriptaddress, 
twocolumn,
prb, 
]{revtex4-2}
\usepackage{graphicx}
\usepackage{color}
\usepackage{hyperref}
\usepackage{CJK}
\begin{document}
\begin{CJK*}{UTF8}{gbsn}

\title{Nonlinear perturbation of a high-order exceptional point: skin discrete breathers and the hierarchical power-law scaling}

\author{Hui Jiang(江慧)}
\thanks{They contribute equally to this work}
\affiliation{School of Physics, South China Normal University, Guangzhou 510006, China}

\author{Enhong Cheng(成恩宏)}
\thanks{They contribute equally to this work}
\affiliation{School of Physics, South China Normal University, Guangzhou 510006, China}

\author{Ziyu Zhou(周子瑜)}
\affiliation{School of Physics, South China Normal University, Guangzhou 510006, China}

\author{Li-Jun Lang (郎利君)}
\email{ljlang@scnu.edu.cn}
\affiliation{School of Physics, South China Normal University, Guangzhou 510006, China}
\affiliation{Guangdong Provincial Key Laboratory of Quantum Engineering and Quantum Materials, School of Physics, South China Normal University, Guangzhou 510006, China}

\date{\today}

\begin{abstract}
We study the nonlinear perturbation of a high-order exceptional point (EP) of the order equal to the system site number $L$ in a Hatano-Nelson model with unidirectional hopping and Kerr nonlinearity. Notably, We find a class of discrete breathers that aggregate to one boundary, here named as {\it skin discrete breathers} (SDBs). The nonlinear spectrum of these SDBs shows a hierarchical power-law scaling near the EP. Specifically, the response of nonlinear energy to the perturbation is given by $E_m\propto \varGamma^{\alpha_{m}}$, where $\alpha_m=3^{m-1}$ is the power with $m=1,\cdots,L$ labeling the nonlinear energy bands. This is in sharp contrast to the $L$-th root of a linear perturbation in general. These SDBs decay in a double-exponential manner, unlike the edge states or skin modes in linear systems, which decay exponentially.
Furthermore, these SDBs can survive over the full range of nonlinearity strength and are continuously connected to the self-trapped states in the limit of large nonlinearity. They are also stable, as confirmed by a defined nonlinear fidelity of an adiabatic evolution from the stability analysis.
As nonreciprocal nonlinear models may be experimentally realized in various platforms, such as the classical platform of optical waveguides, where Kerr nonlinearity is naturally present, and the quantum platform of optical lattices with Bose-Einstein condensates, our analytical results may inspire further exploration of the interplay between nonlinearity and non-Hermiticity, particularly on high-order EPs, and benchmark the relevant simulations.
\end{abstract}

\maketitle
\end{CJK*}

\section{Introduction}
The discrete breathers (DBs), also known as  intrinsic localized modes, that are spatially localized and time-periodic are a class of stable solutions to discrete nonlinear Schr\"{o}dinger equations \cite{EilbeckJohansson2003}, which can exist due to the balance of discreteness and nonlinearity in the underlying systems \cite{FlachGorbach2008}. The discovery of such a solution can dates back to 1969 by Ovchinnikov in one-dimensional chains of coupled anharmonic oscillators \cite{Ovchinnikov1970}, and significant progress has been made in theory since the 1990s \cite{FlachWillis1998}.
Experiments in various platforms \cite{BishopKevrekidis2003,Tsironis2003,Campbell2004,CampbellKivshar2004}, including Josephson-junction arrays \cite{TriasOrlando2000,BinderZolotaryuk2000}, micromechanical arrays \cite{SatoCraighead2004}, optical systems \cite{EisenbergAitchison1998,FleischerChristodoulides2003}, and antiferromagnetic solids \cite{SatoSievers2004}, began to focus on this topic from 1998.

Non-Hermitian physics is a flourishing field \cite{AshidaUeda2020} that has attracted increasing interest in recent years due to the numerous counter-intuitive exotic phenomena without counterparts in Hermitian systems, such as the breaking of parity-time-reversal symmetry \cite{BenderBoettcher1998,GuoChristodoulides2009,PengYang2014,PoliSchomerus2015,LiLuo2019,RenJo2022} and the breakdown of conventional bulk-boundary correspondence \cite{Lee2016,LeykamNori2017,ShenFu2018,YaoWang2018,GongUeda2018,Xiong2018,KunstBergholtz2018,AlvarezTorres2018,YinChen2018,JinSong2019,BorgniaSlager2020,ZhangFang2020}. These phenomena result from the non-Hermiticity of the system's Hamiltonian that gives birth to complex energies and nonunitary evolutions.
Among them, exceptional points (EPs) \cite{Kato1995} are a kind of spectral degeneracy along with the coalescence of corresponding eigenvectors\cite{Heiss2012,AshidaUeda2020,BergholtzKunst2021}.
These unique objects in non-Hermitian systems can induce many exotic phenomena, such as fractional winding numbers \cite{Lee2016,YinChen2018} or mode switching \cite{UzdinMoiseyev2011,BerryUzdin2011,GraefeMoiseyev2013}.

However, DBs have less direct communications with the non-Hermiticity. Some theoretical works \cite{GraefeKorsch2006,GraefeNiederle2010,LangChong2021,Yuce2021,WuZhang2021,RamezanpourBogdanov2021,Ezawa2022a,KhedriZilberberg2022} have already touched upon the crossing scenarios of nonlinearity and non-Hermiticity.
For example, an end breather oscillating at one boundary is proposed in a non-Hermitian Su-Schrieffer-Heeger-like lattice with saturable nonlinearity due to the interplay of non-Hermitian skin effect and the topological end state \cite{LangChong2021}. In one dimensional discrete nonlinear Schr\"{o}dinger equations with nonreciprocal hopping, fractal bands are numerically figured out \cite{Yuce2021}, and several different trapped skin states are formed in quench dynamics \cite{Ezawa2022a}. Nonlinear effects on EPs \cite{KhedriZilberberg2022} or nonlinear EPs \cite{WuZhang2021,RamezanpourBogdanov2021} are also discussed in different non-Hermitian nonlinear models.
In experiments, topological insulator lasers \cite{BahariKante2017,BandresKhajavikhan2018} and the nonlinear Thouless pumping \cite{JurgensenRechtsman2021} are realized by utilizing the merits of chiral edge states of Hermitian topological systems and self-trapping of nonlinearity.

DBs were usually studied in Hermitian nonlinear lattices, i.e., without on-site gain/loss or nonreciprocal hopping that are two forms to break the Hermiticity of systems.
When introducing non-Hermiticity into nonlinear lattices, one may be curious about whether DBs still exist and, if so, whether they are dynamically stable. Further questions arise, such as in what way the breakdown of bulk-boundary correspondence or non-Hermitian skin effect \cite{YaoWang2018} shows up, and whether a nonlinear perturbation induces any exotic scaling around a high-order EP.

In this paper, to address these questions, we introduce an on-site Kerr nonlinearity to a disorder-free Hatano-Nelson model \cite{HatanoNelson1996,HatanoNelson1997}, which is naturally equipped with an EP of a system-size order under open boundary conditions (OBCs) when the nonreciprocal hopping becomes unidirectional.
This gives birth to a class of stable DBs, dubbed {\it skin discrete breathers} (SDBs), which are located near the preferred boundary. These SDBs emerge from the high-order EP and are accompanied by a hierarchical power-law scaling $E_m\propto \varGamma^{\alpha_{m}}$ with $m=1,\cdots,L$ in the power $\alpha_m=3^{m-1}$ labeling the nonlinear energy bands, which reflects the energy response $E$ to the nonlinear perturbation $\varGamma$. This scaling law is in sharp contrast to the linear perturbation. Moreover, these SDBs can stably exist under OBCs through the whole nonlinearity strength and are verified by nonlinear adiabatic evolution. However, under periodic boundary conditions (PBCs), they disappear, showcasing a nonlinear version of the breakdown of bulk-boundary correspondence.
Since nonlinear models with nonreciprocal hopping can be experimentally realized in various platforms, such as optical waveguides \cite{Okamoto2006} where Kerr nonlinearity naturally exists, and optical lattices with Bose-Einstein condensates (BECs) \cite{PethickSmith2008} that can be described by Gross-Pitaevskii (GP) equations, our results may stimulate more studies between the interplay of nonlinearity and non-Hermiticity, especially on high-order EPs.

\section{Model}
To investigate the interplay of nonlinearity and non-Hermiticity, we construct a discrete non-Hermitian nonlinear Schr\"odinger equation with both nonreciprocal hopping and Kerr nonlinearity ($\hbar=1$):
	\begin{equation}
		i\frac{\partial}{\partial t}\varPsi_{n}(t)=J_{\rm L} \varPsi_{n+1}(t)+J_{\rm R}\varPsi_{n-1}(t)+\gamma|\varPsi_n(t)|^2\varPsi_n(t)
		\label{eq:DNSE}
	\end{equation}
	where $\varPsi_n(t)~(n=1,\cdots,L)$ is the time-dependent wave function at the $n$th site of $L$ sites, $J_{\rm L,R}$ generally represent the amplitudes of nonreciprocal hopping if $J_{\rm L}\ne J_{\rm R}$, and $\gamma$ is the strength of Kerr nonlinearity, and here, the parameters $J_{\rm L,R}$ and $\gamma$ are all real numbers.
	This model may be realized in a classical platform of optical waveguides with natural Kerr nonlinearity \cite{JurgensenRechtsman2021} or in a quantum platform with a BEC loaded in a dissipative ultracold atomic system \cite{RenJo2022} (see an example of model derivation using the Lindblad master equation in Appendix \ref{asec:open}).
	
	We are looking for DBs, i.e., temporarily periodic and spatially localized solutions \cite{EilbeckJohansson2003,FlachGorbach2008}, of the form
	\begin{equation}
		\varPsi_{n}(t)=\sqrt{I}\,\psi_ne^{-iE t},
		\label{eq:trial_fun}
	\end{equation}
	where the energy $E$ must be a {\it real} number that conserves the total intensity
	\begin{equation}
		I=\sum_n|\varPsi_{n}(t)|^2,
	\end{equation}
	and $\psi_n$ is the normalized time-independent amplitude, i.e., $\sum_n|\psi_{n}|^2=1$.
	If $E$ is complex, the solution is not time-periodic, i.e., not a DB anymore, and will amplify or decay after a long-time evolution; otherwise, it is time-periodic and possibly stable.
	
	Substituting the trial function Eq. \eqref{eq:trial_fun} into Eq. \eqref{eq:DNSE}, we have
	\begin{equation}
		J_{\rm L}\psi_{n+1}+J_{\rm R} \psi_{n-1}+\gamma I |\psi_n|^2\psi_n=E\psi_{n}.
		\label{eq:static}
	\end{equation}
	For the linear case ($\gamma I=0$), it is just the disorder-free Hatano-Nelson model \cite{HatanoNelson1996,HatanoNelson1997}. Under OBCs, it is well known that $E$ can be a real eigenenergy if and only if $J_{\rm L}J_{\rm R}\ge 0$. Additionally, all the bulk eigenstates assemble at only one boundary if $J_{\rm L}\neq J_{\rm R}$, which is called the non-Hermitian skin effect \cite{YaoWang2018}.
	
	For the nonlinear case ($\gamma \ne0$), we can prove that given a real-valued $\gamma $ as well as the condition $J_{\rm L}J_{\rm R}> 0$, the energy $E$ under OBCs must also be real, as opposed to the case shown in Ref. \cite{Yuce2021}.
	It can be seen as follows: via the transformation of state $\psi_n=t^{n}\phi_n$ with $t=\sqrt{J_{\rm R}/J_{\rm L}}$, Eq. \eqref{eq:static} becomes
	\begin{equation}
		J({\phi}_{n+1}+ {\phi}_{n-1})+\gamma_nI|{\phi}_n|^2{\phi}_n=E{\phi}_{n},
		\label{eq:eqv_mod}
	\end{equation}
	where $J=\sqrt{J_{\rm L}J_{\rm R}}$ and $\gamma_n=t^{2n}\gamma$, and the transformed state satisfies $\sum_nt^{2n}|{\phi}_{n}|^2=1$;
	for any set of solutions $\{{\phi}_n\}$, the coefficient matrix (regarding $|{\phi}_n|^2$ as system parameters) from the left-hand side of Eq. \eqref{eq:eqv_mod} is Hermitian, thus $E$, as the eigenvalue of this Hermitian matrix, must be real.
	Furthermore, because Eq. \eqref{eq:static} is invariant to an overall phase of the solution, we can always make $\psi_1$ nonnegative, and thus, all other $\psi_n$'s must be real, which is apparent from the recurrence relation in Eq. \eqref{eq:static} with the above proved reality of $E$.
	Therefore, the solution $\psi_n$ to Eq. \eqref{eq:static} just needs to be found in the field of real numbers.
	
	According to the above discussion, we can just consider the cases of $0\le J_{\rm R}/J_{\rm L}\le 1$ with $J_{\rm L}=1$ being set as the energy unit in the following; here, we also include the reciprocal case of $J_{\rm R}=1$ and the unidirectional case of $J_{\rm R}=0$.
	For convenience, we define an effective nonlinear strength $\varGamma\equiv\gamma I$ (where, in contrast, $\gamma$ is called the bare strength) of nonlinearity, which will be set positive afterward, and the negative case can be directly obtained by the mapping: $(\varGamma,E,\psi_n)\rightarrow(-\varGamma,-E,(-1)^n\psi_n)$.
	
	For the unidirectional case of $J_{\rm R}=0$, it is well known that under OBCs the linear limit (i.e., $\varGamma=0$) corresponds to an exceptional point of order $L$ with the eigenenergy $E=0$ and the unique eigenstate $(1,0,0,\cdots)^{\rm T}$. Then, a question arises:  how does Kerr nonlinearity affect this high-order exceptional point?
	In the following, we mainly pay attention to this case and then discuss the general nonreciprocal case in the end.

\begin{figure}[tb] 
    \centering
	\includegraphics[width=1\linewidth]{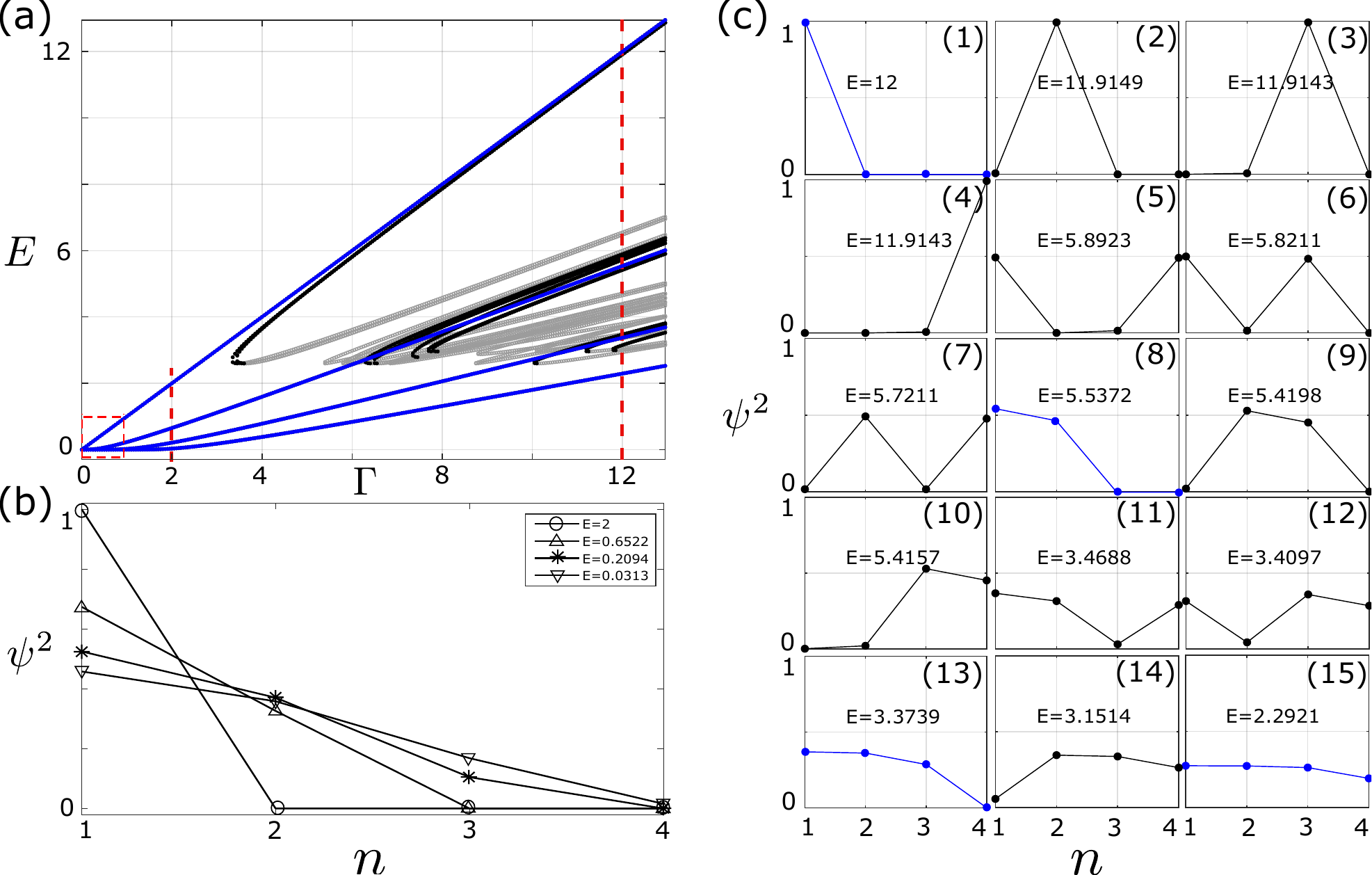}
    \caption {(a) Nonlinear energy spectrum, $E$ versus $\varGamma$, under OBCs with $L=4$ sites by the numerical method in Appendix \ref{asec:method}. 
    The four blue curves ($m=1,2,3,4$ from top to bottom) emerging from the linear EP of order 4 are just SDBs.
    Stable DBs are represented by blue or black dots while the unstable ones are by gray circles, calculated using
    Eq. \eqref{eq:stable} with accuracy $\tau=10^{-16}$.
	(b,c) Intensity profiles of stable DBs at relatively small and large nonlinearities, $\varGamma=2$ and $12$,
    respectively, indicated by the red dashed line in (a). The intensity profiles colored in blue (c1,c8,c13,c15) correspond to the SDBs in (a).}\label{fig1}
\end{figure}

	Although the previous proof of the reality of solutions is only applicable to the $J_{\rm R}>0$ case, we can believe that it is still true for $J_{\rm R}\rightarrow 0$, or at least, we can still only care about the real solutions, i.e., $\{\psi_n, E\}\in \mathbb{R}$.
For the unidirectional case, Eq. \eqref{eq:static} reduces to
\begin{equation}
	\psi_{n+1}+\varGamma \psi_n^3=E\psi_{n},
	\label{eq:one-way}
\end{equation}
with the normalization condition $\sum_n\psi_n^2=1$.
In Fig. \ref{fig1}, we numerically plot the nonlinear energy spectrum under OBCs with $L=4$ sites, and show typical intensity profiles $\psi_n^2$ at relatively small and large nonlinearities. The numerical method used here can be referred to in Appendix \ref{asec:method}.

\section{Hierarchical power-law scaling and self-trapping}
	Near the linear high-order EP at $\varGamma=0$, it is shown that the perturbation by Kerr nonlinearity $\varGamma\ll 1$ lifts the exceptional degeneracy of order $L$ to $L$ nonlinear energies [Fig. \ref{fig1}(a)], indicating a hierarchy of power-law scalings when $\varGamma\rightarrow 0$.
	Correspondingly, the wave functions of nonlinearly perturbed energies are just located to the preferred left boundary with tails extending to interior sites, dubbed {\it skin discrete breathers} (SDBs). The lower the energy is, the farther the tail extends away from the boundary, as shown in Fig. \ref{fig1}(b).
	
	To exactly find the power-law scaling when $\varGamma\ll1$, we try to solve Eq. \eqref{eq:one-way} with the assumption of $\psi_m\ne 0$ and $\psi_{n> m}=0$, inspired by the intensity profiles of wave functions at weak nonlinearity in Fig. \ref{fig1}(b). Thus, Eq. \eqref{eq:one-way} can be rewritten in a form of nonlinear recurrence relations, terminated at the $m$-th site, i.e.,
	\begin{equation}
		\left\{
		\begin{aligned}
			\psi_{n+1}&=(E-\varGamma \psi_n^2)\psi_n, &(1\le n< m)\\
			0&=(E-\varGamma \psi_m^2)\psi_m.&\\
		\end{aligned}
		\right.
		\label{eq:recurrence}
	\end{equation}
	Since we care about the perturbation to the linear high-order EP at $\varGamma=E=0$, we can reduce the relation \eqref{eq:recurrence} using the perturbation condition of $\{\varGamma,E\}\ll 1$.
	The basic idea is to retain only the terms independent of $E$ in each recursion step of the first line of Eq. \eqref{eq:recurrence}, and then to solve $E$ through the second line.
	The details of derivation can be referred to in Appendix \ref{asec:scaling}.
	
	As the result of derivation, we get the nonlinearly perturbed energy $E_m(\varGamma)$ of the SDB that spreads maximally into site $m$,
	\begin{equation}
		E_m(\varGamma)\approx \varGamma^{\alpha_{m}},
		\label{eq:energy}
	\end{equation}
	and the corresponding (unnormalized) wave function,
	\begin{eqnarray}
		\psi_n(\varGamma)&\approx&(-\varGamma)^{\frac{\alpha_{n}-1}{2}} ~~~(n=1,\cdots m),
		\label{eq:wavefunction}
	\end{eqnarray}
	where $\alpha_n=3^{n-1}$.
	A power-law scaling emerges in the dependence of both the energy and the wave function on $\varGamma$.
	
	From Eq. \eqref{eq:energy}, it is shown that the nonlinear perturbation indeed lifts the linear EP of order $L$ to $L$ nonlinear energies $E_m~(m=1,\cdots,L)$, as expected from Fig.\ref{fig1}(a), which is similar to a linear perturbation.
It is well known that a linear perturbation $\delta$ will in general lead to a spectral shift $\epsilon\propto\sqrt[L]{\delta}$, according to the Puiseux's expansion \cite{Kato1995}.
	Although our nonlinear perturbation $\varGamma$ also gives rise to a power-law scaling $\epsilon\propto \varGamma^{\alpha_m}$, the power $\alpha_m$ {\it exponentially} depends on $m$, i.e., $\alpha_m=3^{m-1}$, where the base 3 is inherited from the power of Kerr nonlinearity, and demonstrates a hierarchical structure for $m=1,\cdots,L$. This is in sharp contrast to the uniform fractional power $1/L$ of the linearly perturbed energies.
	From the sensitivity point of view \cite{AshidaUeda2020}, the energy response to the nonlinear perturbation is more insensitive than to the linear one.
	
	From Eq. \eqref{eq:wavefunction}, we have $\psi_{n+1}(\varGamma)/\psi_n(\varGamma)=(-\varGamma)^{\alpha_{n}}$, which means that the wave function of the $m$-th energy decays exponentially into the interior sites, but the power $\alpha_n=3^{n-1}$ is much distinct to the power $n$ of a general linear end state, indicating a faster decay.
	Note that $m$ also labels the farthest site the DB of energy $E_m(\varGamma)$ occupies, representing DBs with different distributions. From the previous discussion of sensitivity, this means that the distribution of DBs can influence the energy response to the nonlinear perturbation: the more sites the DB occupies, the more insensitive the energy responses to the perturbation.
	
\begin{figure}[tb] 
	\includegraphics[width=1\linewidth]{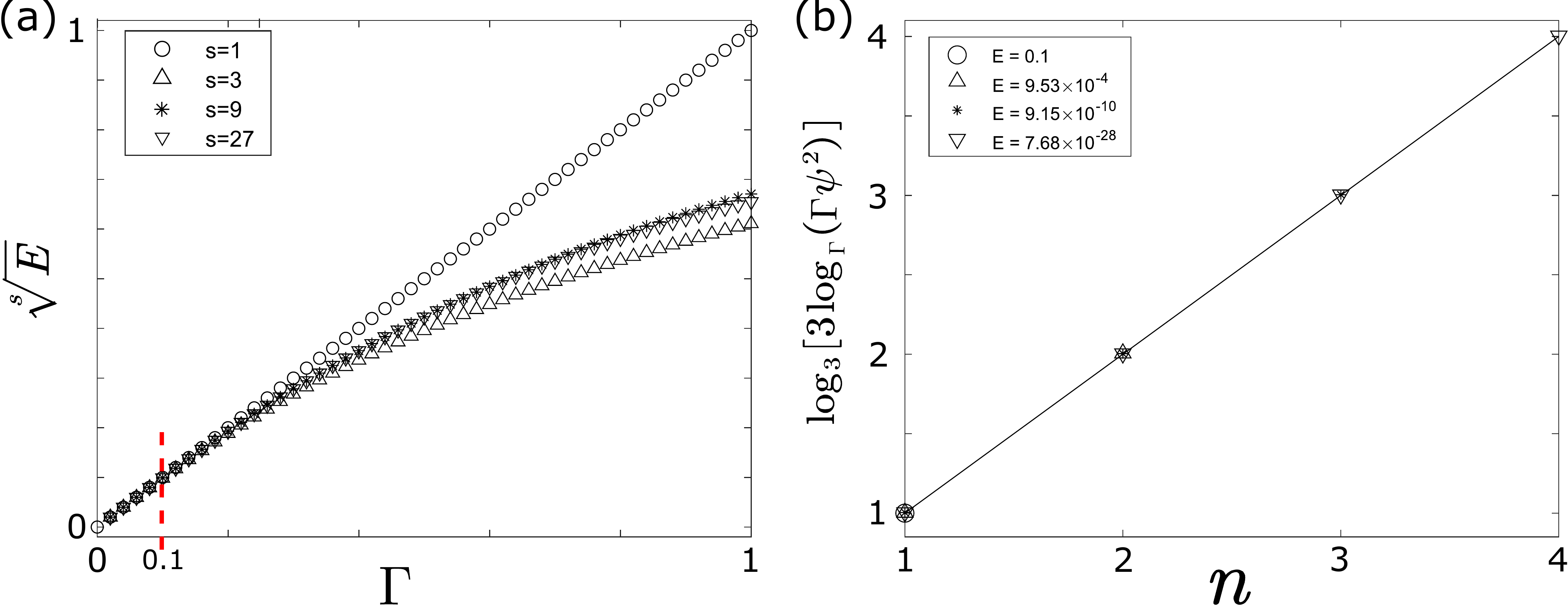}\\
    \centering
    \caption{(a) The nonlinear energies in the red dashed rectangle in Fig.\ref{fig1}(a), replotted as $\sqrt[s]{E}~(s\equiv\alpha_m=1,3,9,27$ for $m=1,2,3,4$, respectively) versus $\varGamma$. They are almost linear with slope one for small $\varGamma$, satisfying Eq. \eqref{eq:energy}.
    (b) The wave functions of the four energies at $\varGamma=0.1$, indicated by the red dashed vertical line in (a),
    replotted as  $\log_3[3\log_\varGamma(\varGamma\psi^2)]$ versus the site index $n=1,\cdots,m$, where $m=1,2,3,4$ correspond to the four energies in the decreasing order. The numerical results labeled by markers coincide with the solid line of slope one, satisfying Eq. \eqref{eq:wavefunction}.}\label{fig2}
\end{figure}

	These scaling laws in energies and wave functions can be verified for $L=4$, as shown in Fig. \ref{fig2}.
	It is shown that when $\varGamma\rightarrow 0$, the nonlinear energies numerically obtained in Fig. \ref{fig1}(a) indeed have the hierarchical power-law scalings $E_m(\varGamma)\propto \{\varGamma, \varGamma^{3}, \varGamma^{9}, \varGamma^{27}\}$ from top to bottom corresponding to $m=\{1,2, 3,4\}$, respectively [Fig. \ref{fig2}(a)], and the corresponding wave functions decay in a double-exponential way from the left boundary (site 1) to site $m$ [Fig. \ref{fig2}(b)], satisfying the analytical results from Eqs. \eqref{eq:energy} and \eqref{eq:wavefunction}.
	
When the nonlinearity is strong enough (i.e., $\varGamma\rightarrow \infty$), the self-trapping is expected to occur. In this limit, Eq. \eqref{eq:one-way} can be decoupled as
\begin{equation}
	E\psi_{n}\approx \varGamma \psi_n^3,
\end{equation}
with the normalization condition $\sum_{n=1}^{L}\psi_n^2=1$.
From this equation, we know that if there is an occupation at site $n$, the energy satisfies $E\approx \varGamma\psi_n^2$, which means that all occupied sites must have the same intensity $\psi_{n\in {\mathbb O}}^2\approx E/\varGamma$ with ${\mathbb O}\equiv\{n|\psi_n\ne 0, 1\le n\le L,n\in\mathbb{N}\}$ being the set of all occupied sites.
Therefore, using the normalization condition, we have the energy,
\begin{equation}
	E_m\approx \varGamma/m,
\end{equation}
and the corresponding wave function,
\begin{equation}
	\psi_{n\in {\mathbb O}}^2\approx 1/m, ~~~ \psi_{n\not\in {\mathbb O}}^2\approx 0,
\end{equation}
where $m$ is the number of occupied sites, i.e., the self-trapping occurs at $m$ independent sites.
From the possible arrangement of $m$ sites being occupied, the energy $E_m$ has the degeneracy of $C_L^m$, and at large $\varGamma$ there exist $L$ separated bands, each of which includes an SDB adiabatically evolved from the linear EP, as shown in Fig. \ref{fig1}(a).
This self-trapping seems similar to the Hermitian nonlinear case, but for a large but finite strength of nonlinearity, the perturbation of unidirectional hopping leads the trapping tail spreading asymmetrically to the left side of the trapped sites, not as symmetric as its Hermitian counterpart.

\begin{figure}[tb] 
	\includegraphics[width=1\linewidth]{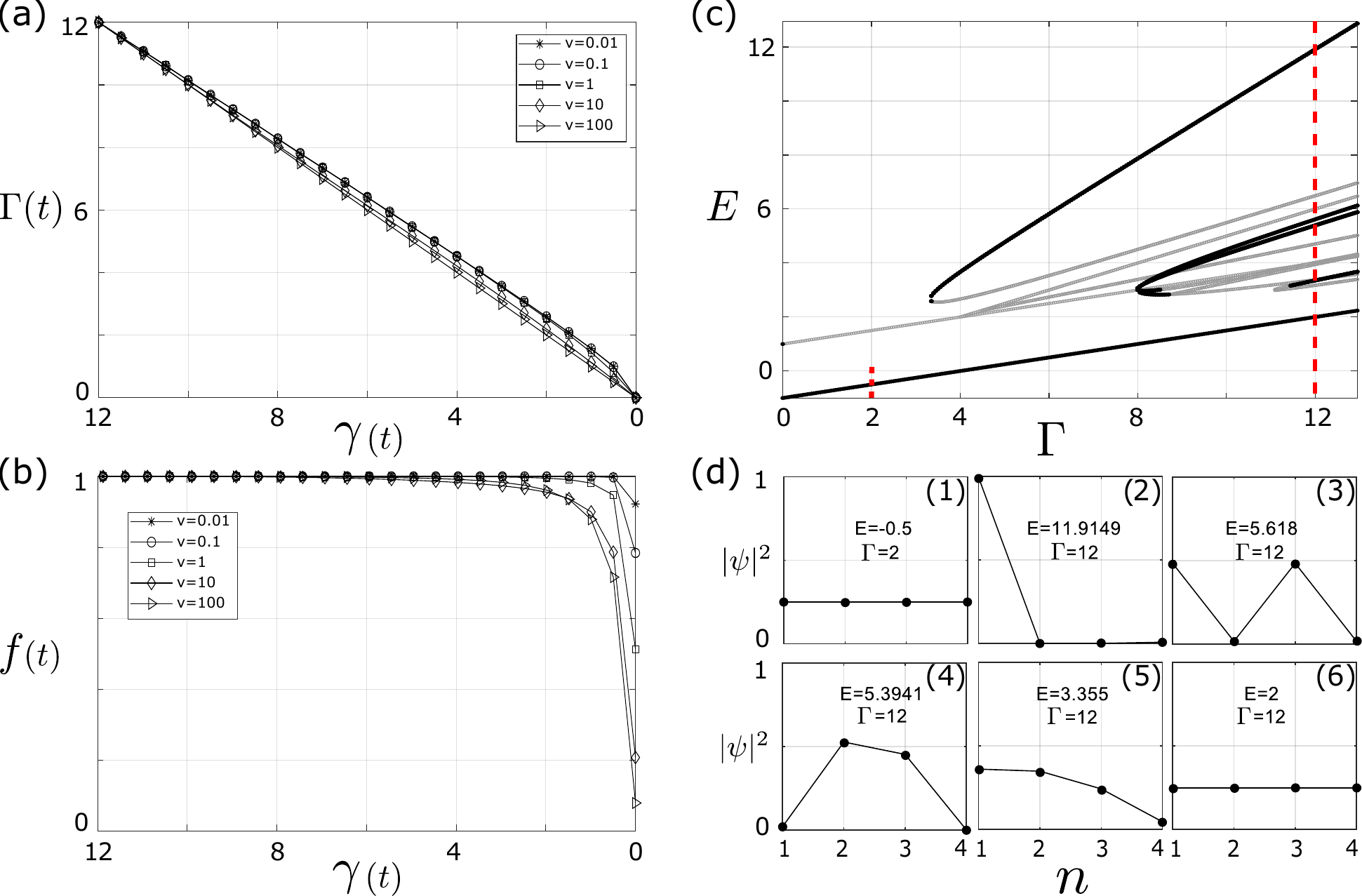}\\
    \centering
	\caption{
    (a) The instantaneous effective strength $\varGamma(t)=\gamma(t)I(t)$ versus the bare strength $\gamma(t)=\gamma_0-vt$, indicating the nonuniform change of $\varGamma(t)$ due to the increase of intensity $I(t)$ when the nonreciprocal hopping gradually takes part in along with time $t$. The smaller speed $v$ leads to larger nonuniformity of $\varGamma(t)$'s change.
    (b) The nonlinear fidelity of the instantaneous state to the SDB at $\varGamma(t)$ along the energy curve of $m=2$ in Fig.\ref{fig1}(a), which shows that the nonlinear adiabaticity works well if the change is slow enough.
    In (a) and (b), the initial SDB state is selected as in Fig.\ref{fig1}(c8) of 2 leftmost sites being occupied (i.e., $m=2$) at $\varGamma(t=0)=\gamma_0=12$.
    (c) Nonlinear energy spectrum with the same settings as in Fig.\ref{fig1}(a) except under PBCs, where the energies can be complex in general but only real ones are plotted, including stable (black dots) and unstable (gray hollow circles) solutions.
    (d) Some typical intensity profiles of stable solutions in (c), the lattice translation of each of which is also the degenerate solution due to the translation invariance under PBCs.}\label{fig3}
\end{figure}

\section{Stable skin breathers and nonlinear adiabaticity}
As shown in Fig. \ref{fig1}(a) that the numbers of energies at these two limits (i.e., $\varGamma\rightarrow 0$ and $\infty$) are different, there must be emergent states at intermediate $\varGamma$.
Among them, we only care about the dynamically stable states.

We first analyze the linear stability of amplitude \cite{CarrEilbeck1985,MacKaySepulchre1998,JurgensenRechtsman2021}, i.e., consider the perturbation $u_n(t)$ on the amplitude of the DB solution $\varPsi_n(t)=\psi_ne^{-iEt}$, where $(\psi_n,E)$ is the solution to Eq. \eqref{eq:one-way}.
The perturbed DB, say $\tilde{\varPsi}_n(t)\equiv[\psi_n+u_n(t)]e^{-iEt}$, must satisfy the nonlinear time-dependent Schr\"{o}dinger equation,
\begin{eqnarray}
	i\frac{\partial}{\partial t}\tilde{\varPsi}_{n}(t)=\tilde{\varPsi}_{n+1}(t)+\varGamma|\tilde{\varPsi}_n(t)|^2\tilde{\varPsi}_n(t),
\end{eqnarray}
yielding [up to the first order of $u_n(t)$]
\begin{equation}
	\left\{
	\begin{aligned}
	\partial_t{u}^{\rm (r)}_{n}(t)&\approx (\varGamma\psi_n^2-E)u_n^{\rm(i)}(t)+u_{n+1}^{\rm(i)}(t) \\
	\partial_t{u}^{\rm(i)}_{n}(t)&\approx -(3\varGamma\psi_n^2-E)u_n^{\rm(r)}(t)-u_{n+1}^{\rm(r)}(t)
\end{aligned}
\right.,
\label{eq:stability}
\end{equation}
where the subscripts ${\rm(r/i)}$ represent the real and imaginary parts of $u_n(t)$.
In matrix form, it can be written as
\begin{equation}
	\frac{\partial}{\partial t}
	\left[
	\begin{matrix}
		{U}^{\rm(r)}(t)\\
		{U}^{\rm(i)}(t)
	\end{matrix}
\right]
=
	\mathcal{C}\left[
\begin{matrix}
	{U}^{\rm(r)}(t)\\
	{U}^{\rm(i)}(t)
\end{matrix}
\right],
\end{equation}
where ${U}^{\rm(r,i)}(t)=[u^{\rm(r,i)}_1(t),\cdots,u^{\rm(r,i)}_L(t)]^{\rm T}$ and $\mathcal{C}$ is the corresponding $2L\times 2L$ coefficient matrix, determined by the given solution $(\psi_n,E)$ to Eq. \eqref{eq:one-way} for each $\varGamma$.
The stable states require that $u_n(t)$ for all $n$ must not be amplified during the evolution, in other words, the real parts of all the eigenvalues $\{\lambda_n\}~(n=1,\cdots,2L)$ of $\mathcal{C}$ must not greater than $0$, i.e.,
\begin{eqnarray}
	\max[\text{Re}(\lambda_n)]\le\tau\sim0,
	\label{eq:stable}
\end{eqnarray}
where $tau$ is the accuracy taken in numerical calculations.

Therefore, we can numerically distinguish the stable and unstable solutions, labeled by dots and hollow circles, respectively, in Fig. \ref{fig1}(a), where the SDBs near the linear high-order EP at weak nonlinearity are all found stable, and each of them with energy $E_m$ can go through the full range of $\varGamma$ to a boundary-$m$-site-occupied self-trapped SDB at strong nonlinearity [Fig. \ref{fig1}(b)].
However, other stable DBs localized at the interior sites emerge at finite $\varGamma$.
For example, the DB localized at site 2 [Fig. \ref{fig1}(c2)] does not emerge from the EP at $\varGamma=0$ but from a finite $\varGamma\approx 3.33$ (see detailed calculation in Appendix \ref{asec:method}). 
This is similar to the commonly encountered bifurcation point in Hermitian nonlinear systems where the self-trapping bifurcates from a continuous spectrum of extended states. 
Here, it emerges from a bunch of skin modes, which themselves can also be regarded as the boundary self-trapped modes.

Since these SDBs are stable, we may check their adiabaticity along with the change of bare strength of nonlinearity $\gamma(t)=\gamma_0-vt$ in time $t$, where $\gamma_0$ is the initial strength and $v$ is the change speed.
As noted that the intensity $I(t)=\sum_n|\varPsi_n(t)|^2$ of the state $\varPsi_n(t)$ governed by Eq. \eqref{eq:DNSE} generally changes in time due to the system's non-Hermiticity, we need to compare the state with the instantaneous SDB $\psi_n[\varGamma(t)]$ for $\varGamma(t)\equiv\gamma(t)I(t)$.
To reflect the robustness of the adiabatic process, we define a nonlinear fidelity,
\begin{eqnarray}
	f(t)=\frac{1}{I(t)}\left|\sum_n\varPsi_n(t)\psi_n[\varGamma(t)]\right|^2.
\end{eqnarray}
The numerical results in Fig. \ref{fig3}(b) show the robustness of SDBs: the slower the $\gamma(t)$ changes, the more similar the instantaneous state is to the SDB at $\varGamma(t)$. Near the linear EP, the adiabaticity tends to break down, which may result from the divergence of the state amplitude near EP \cite{CM}.
Because of the dependence of $\varGamma(t)$ on the increasing intensity $I(t)$ along with time, the nonlinear adiabatic evolution is not uniform along a nonlinear energy curve of an SDB [Fig. \ref{fig3}(b)], which is different from the linear adiabatic evolution.

As a comparison, we also plot the nonlinear real spectrum as well as the typical intensity profiles under PBCs in Figs. \ref{fig3}(c) and \ref{fig3}(d). Note that complex solutions to Eq. \eqref{eq:one-way} exist under PBCs, and here only real ones are plotted for the sake of dynamical stability.
Compared with the spectrum under OBCs in Fig. \ref{fig1}(a), the SDBs are missing except for the all-site-occupied state. The breakdown of spectrum correspondence in non-Hermitian linear systems between OBCs and PBCs also occurs in this non-Hermitian nonlinear system.
Besides the similar self-trapped DBs at interior sites, SDBs, i.e., the DBs aggregated to one boundary, emerge under OBCs.

\begin{figure}[tb] 
	\includegraphics[width=1\linewidth]{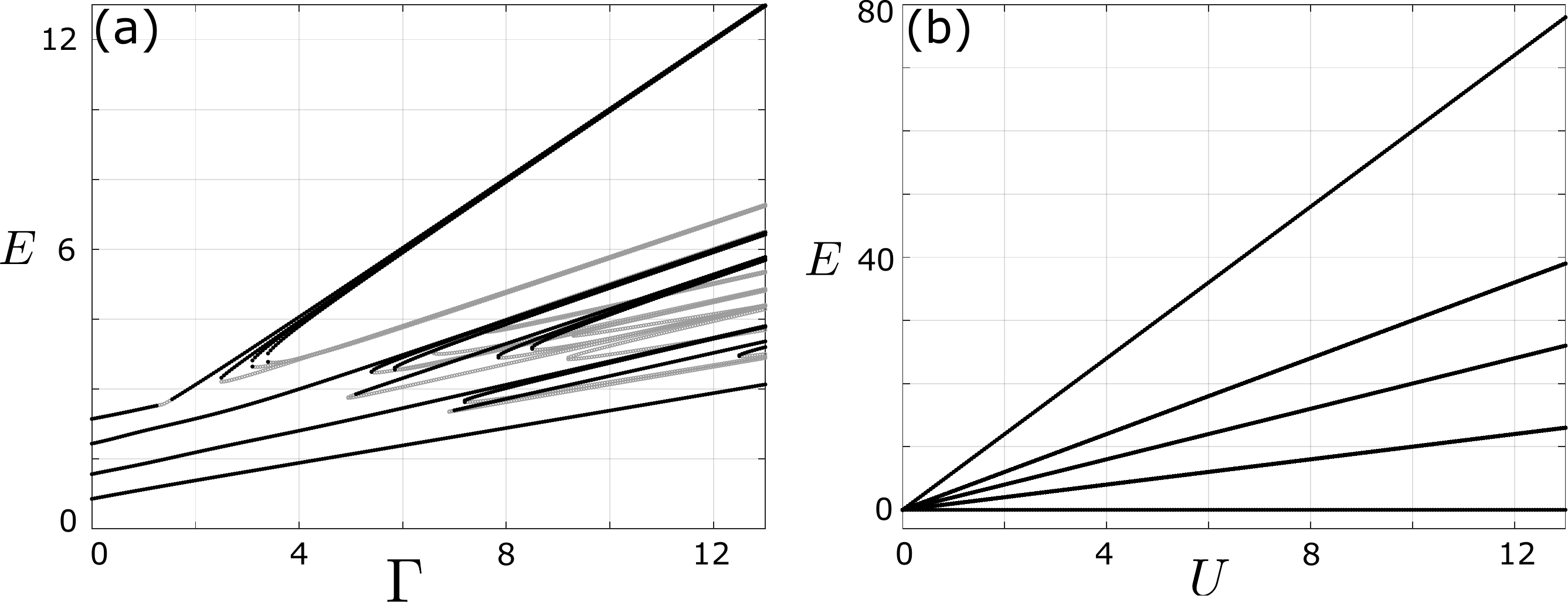}\\
    \centering
	\caption{
    (a) Nonlinear energy spectra with the same setting as in Fig. \ref{fig1}(a) but with the nonreciprocal hopping $J_{\rm R}/J_{\rm L}=0.5$.
    (b) The energy spectrum of the many-body Hamiltonian \eqref{eq:many} with $L=4$ sites and $N=4$ bosons under OBCs, where the particle distribution $\langle\hat{n}_i\rangle\equiv\langle \psi_m |\hat{n}_i|\psi_m\rangle$ with $|\psi_m\rangle$ being the right eigenstate of $\hat{H}_{\rm MB}$ normalized to $N=4$, corresponding to the $m$-th energy band from top to bottom. Here $J_{\rm L}=1$ is set as the energy unit.}
	\label{fig4}
\end{figure}

\section{Discussion and conclusion}
Although we only demonstrate the unidirectional case of $J_{\rm R}=0$, the nonreciprocal case with $J_{\rm R}/J_{\rm L}>0$ can be studied numerically in the same way, which shows that the hierarchical scaling law disappears along with the EP [Figs. \ref{fig4}(a)].
Since Kerr nonlinearity sometimes can be regarded as a mean-field approximation of some many-body systems (see Appendix \ref{asec:open} for an example), here we also plot the energy spectrum [Fig. \ref{fig4}(b)] of a repulsive Bose-Hubbard model with only the leftward hopping, of which the Hamiltonian reads
\begin{eqnarray}
	\hat{H}_{\rm MB}=-J_{\rm L}\sum_{i=1}^{L-1}\hat{a}_i^\dag\hat{a}_{i+1}+\frac{U}{2}\sum_{i=1}^L\hat{n}_i(\hat{n}_i-1),
	\label{eq:many}
\end{eqnarray}
where $\hat{a}_i^{(\dag)}$ is the bosonic annihilation (creation) operator at the $i$th site, $\hat{n}_i=\hat{a}_i^{\dag}\hat{a}_i$ the corresponding particle number operator, $J_{\rm L}$ the strength of the leftward hopping, and $U>0$ the two-body interaction strength.
It is shown that there is no such scaling law in the spectrum near the high-order EP.
The connection between this linear many-body interacting model to the nonlinear one in the main part should be further investigated.
	
In conclusion, we study a nonlinear model with nonreciprocal hopping, especially with unidirectional hopping which corresponds to a high-order EP in the linear limit. By analyzing the nonlinear energy spectrum,  we find a class of SDBs under OBCs, inherited from the skin modes of the linear systems. At weak nonlinearity, the linear EP gives rise to a hierarchical power-law scaling induced by the nonlinear perturbation. These SDBs can continuously extend to the large intensity regime, becoming boundary self-trapped states.
These SDBs are dynamically stable, and the nonlinear adiabaticity is still valid as long as the nonlinearity strength changes slowly enough.
The SDBs and relevant phenomena in nonreciprocal nonlinear models could be experimentally observed in many platforms, such as optical waveguides, where Kerr nonlinearity naturally exists, and optical lattices with BECs, which can be described by GP equations.

\section*{Acknowledgment}
This work was supported by the National Key Research and Development Program of China (Grant No.~2022YFA1405304), the Key-Area Research and Development Program of Guangdong Province (Grant No.~2019B030330001), and the Guangdong Provincial Key Laboratory (Grant No.~2020B1212060066).


\appendix
\section{A possible realization of model {\eqref{eq:DNSE}} with ultracold atoms}\label{asec:open}
Considering a BEC formed by weakly repulsive ultracold bosons loaded in a dissipative optical lattice, under the Markov approximation \cite{BreuerPetruccione2002}, the Lindblad master equation can be engineered as follows
\begin{eqnarray}
	\frac{d\hat{\rho}}{dt}=-i[\hat{H},\hat{\rho}]-\frac{1}{2}\sum_i(\hat{L}_i^\dag\hat{L}_i\hat{\rho}+\hat{\rho}\hat{L}_i^\dag\hat{L}_i-2\hat{L}_i\hat{\rho}\hat{L}_i^\dag),\notag\\
	\label{eq-lindblad}
\end{eqnarray}
where $\hat{\rho}$ is the density operator, and
\begin{eqnarray}
	\hat{H}=\sum_{i}\left[-J(\hat{a}_i^\dag\hat{a}_{i+1}+\hat{a}_{i+1}^\dag\hat{a}_{i})+\frac{U}{2}\hat{n}_i(\hat{n}_i-1)\right]
\end{eqnarray}
is the conventional Bose-Hubbard Hamiltonian with $\hat{a}_i^{(\dag)}$ being the bosonic annihilation (creation) operator at the $i$-th site, $\hat{n}_i=\hat{a}_i^{\dag}\hat{a}_i$ the corresponding particle number operator, $J$ the hopping strength, and $U>0$ the two-body repulsively contact interacting strength.
Here, we consider a kind of non-local single-particle loss represented by the Lindblad operators \cite{SongWang2019},
\begin{eqnarray}
	\hat{L}_i=\sqrt{g}(\hat{a}_i-i\hat{a}_{i+1}),
\end{eqnarray}
with $g$ being the loss rate.

We can rewrite Eq. \eqref{eq-lindblad} in terms of an effective non-Hermitian Hamiltonian
\begin{eqnarray}
	\hat{H}_{\rm eff}=\hat{H}-\frac{i}{2}\sum_i\hat{L}_i^\dag\hat{L}_i,
\end{eqnarray}
yielding
\begin{eqnarray}
	\frac{d\hat{\rho}}{dt}=-i(\hat{H}_{\rm eff}\hat{\rho}-\hat{\rho}\hat{H}_{\rm eff}^\dag)+\sum_i\hat{L}_i\hat{\rho}\hat{L}_i^\dag, \label{eq:Lindblad-eff}
\end{eqnarray}
where the last term is the recycling term \cite{Daley2014} that can be ignored for a short-time evolution before the quantum jumps occur.
Thus, the dynamics can be well described by the effective non-Hermitian Hamiltonian $\hat{H}_{\rm eff}$, of which the explicit form is
\begin{eqnarray}
	\notag
 	\hat{H}_{\rm eff}&=&
 \sum_{i}\big[-(J+\frac{g}{2})\hat{a}_i^\dag\hat{a}_{i+1}-(J-\frac{g}{2})\hat{a}_{i+1}^\dag\hat{a}_{i}\\ 
 &-&ig\hat{n}_{i}
 +\frac{U}{2}\hat{n}_i(\hat{n}_i-1)\big].
\end{eqnarray}

In the following, we derive the discrete non-Hermitian nonlinear Schr\"odinger equation (\ref{eq:DNSE}) by calculating the dynamics of the single-particle reduced density matrix, $\varDelta_{ij}\equiv\langle\hat{a}_i^\dag\hat{a}_i\rangle={\rm tr}(\hat{a}_i^\dag\hat{a}_i\hat{\rho})$, under mean-field approximation  \cite{WitthautWimberger2011}.
From Eq. (\ref{eq:Lindblad-eff}), we have
 \begin{eqnarray}
 		\notag
		i\frac{d\varDelta _{ij}}{dt}&=& \left( J-\frac{g}{2} \right) \varDelta _{i-1,j}-\left( J-\frac{g}{2} \right) \varDelta _{i,j-1}\\  
		&+&\left( J+\frac{g}{2} \right) \varDelta _{i+1,j}-\left( J+\frac{g}{2} \right) \varDelta _{i,j+1}\\ \notag
		&-&2ig \varDelta _{ij}+U\left(\sigma_{ijjj}+\varDelta _{ij}\varDelta _{jj}-\sigma _{iiij}-\varDelta _{ii}\varDelta _{ij} \right),
		\label{eq:single-density}
\end{eqnarray}
where we have defined the covariances
 \begin{eqnarray}
	\sigma_{ijkl}=\langle a_{i}^{\dagger}a_ja_{k}^{\dagger}a_l \rangle -\langle a_{i}^{\dagger}a_j \rangle \langle a_{k}^{\dagger}a_l \rangle.
\end{eqnarray}
Under the mean-field approximation, we can remove the covariance terms in Eq. (\ref{eq:single-density}) to make this set of equations closed:
 \begin{eqnarray}
	i\frac{d}{dt}\varDelta _{ij}&=&
	\left( J-\frac{g}{2} \right) \varDelta _{i-1,j}-\left( J-\frac{g}{2} \right) \varDelta _{i,j-1} \notag\\
	&+&\left( J+\frac{g}{2} \right) \varDelta _{i+1,j}-\left( J+\frac{g}{2} \right) \varDelta _{i,j+1} \notag\\
	&+&~U\left(\varDelta _{ij}\varDelta _{jj}-\varDelta _{ii}\varDelta _{ij} \right)-2ig \varDelta _{ij}.
	\label{eq:single-density-close}
\end{eqnarray}
This assumption is appropriate for a pure BEC \cite{WitthautWimberger2011}.

By the identification $\varDelta_{ij}= \varPsi_i^*\varPsi_j$ with $\varPsi_i$ being the BEC's wave function at the $i$-th site, Eq. (\ref{eq:single-density-close}) can be re-expressed as
 \begin{equation}
    i\frac{d}{dt}\varPsi_{i}=
    -\left( J-\frac{g}{2} \right) \varPsi _{i-1}-\left( J+\frac{g}{2} \right) \varPsi_{i+1} +U|\varPsi _{i}|^2\varPsi _i -ig \varPsi _i,
\end{equation}
which is just the discrete non-Hermitian nonlinear Schr\"odinger equation (\ref{eq:DNSE}) up to an overall loss term $-ig \varPsi _i$; the interacting strength $U$ takes the role of the strength of the Kerr nonlinearity and the loss rate $g$ controls the nonreciprocal hopping $J_{\rm L,R}\equiv-(J\pm g/2)$.
By fine-tuning the loss rate, one can make the hopping unidirectional, i.e., $g=2J$.

\section{Analytical and numerical methods for nonlinear energy spectra}\label{asec:method}
Indicated by the intensity profiles of the SDBs at weak nonlinearity in Fig.\ref{fig1}(b) of the main text, one can rewrite Eq. \eqref{eq:recurrence} of the main text in a recurrence form, terminated at the $m$-th site, i.e.,
	\begin{equation}
	\left\{
	\begin{aligned}
		\psi_{n+1}&=(E-\varGamma \psi_n^2)\psi_n &(1\le n< m)\\
		0&=(E-\varGamma \psi_m^2)\psi_m&\\
		0&=\psi_n&(n>m)
	\end{aligned}
	\right..
	\label{aeq:recurrence}
\end{equation}
	
	If the DB is localized at the leftmost site, i.e., site 1,  according to the assumption that $\psi_2=0$, one can easily obtain that $\psi_1=\sqrt{E/\varGamma}$ from Eq. \eqref{aeq:recurrence}.
	Further considering the normalization condition $\psi_1^2=1$, one can immediately find that the energy $E_1=\varGamma$ and the corresponding DB solution localized at site 1 is $\varPsi_1=(1,0,0,\cdots)^{\rm T}$.
	This boundary DB is identical to the skin mode of the linear case (i.e., $\varGamma=0$) but its energy becomes finite and proportional to $\varGamma$ due to Kerr nonlinearity.
	
	Likewise, if the DB is localized at site 2, one can determine that $\psi_3=0$, $\psi_2=\sqrt{E/\varGamma}$, and $\psi_1=\sqrt{1-E/\varGamma}$.
	Because the real amplitude $\psi_2<1$,  the energy of the DB localized away from the boundary will smaller than the DB at the boundary, i.e.,  $E_2<\varGamma=E_1$,  which is due to the repulsion of the peak intensity from one site to more sites to the left, lowering the total energy.
	From Eq. \eqref{aeq:recurrence}, the energy $E_2$ is determined by
	\begin{equation}
		4E_2^3-8\varGamma E_2^2+(5\varGamma^2+1)E_2-\varGamma^3 =0.
		\label{aeq:site2}
	\end{equation}
	Using the discriminant \cite{Dickson1914},
	\begin{eqnarray}
		\varDelta&=&18abcd-4b^3d+b^2c^2-4ac^3-27a^2d^2 \notag\\
		&=&16(\varGamma^4-11 \varGamma^2-1),
		\label{aeq:discriminant}
	\end{eqnarray}
 of a cubic equation, $ax^3+bx^2+cx+d=0~(a\ne 0)$, the bifurcation point,  $\varGamma_c=\sqrt{(11+5\sqrt{5})/2}\approx 3.33$, is obtained by $\varDelta=0$, which separates regions of one and three real solutions of Eq. \eqref{aeq:site2} and is also the onset of the DB localized at site 2 as shown in Fig. \ref{fig1}(c2) of the main text.
	
The DB localized at site $n~(n\ge 3)$ is hard to solve analytically due to the nonexistence of  analytical expression of higher-order polynomial equations in $E$. Thus, one may resort to numerical methods to get the full nonlinear energy spectrum as well as the corresponding wave functions, and maybe also the bifurcation points of various DBs localized at interior sites. The plots in Fig. \ref{fig1}(a) of the main text are obtained using the build-in function NSolve over the real field in Mathematica to deal with the system of Eq. \eqref{aeq:recurrence}.
Especially for very small $\varGamma$'s, we restrict the solutions to have nonzero amplitudes before the $m$-th site, i.e., $\psi_{n\le m}>0$, obtaining each curve of $E_m$ versus $\varGamma$ with high numerical accuracy up to 100 significant digits set by the build-in function SetPrecision in Mathematica.
	
\section{Detailed derivation of the power-law scaling of SDBs}\label{asec:scaling}
	Here, we give the detailed derivation of the scaling law at weak Kerr nonlinearity, i.e., $\varGamma\ll 1$, near the linear high-order EP ($\varGamma=E=0$) in the main text.
	To this end, we revisit the recurrence relations Eq. \eqref{aeq:recurrence} and pay attention to the solution $[E(\varGamma),\psi_{n}(\varGamma)]$ when $\varGamma\ll 1$.
	
	Given the amplitude $\psi_1(\varGamma)$ at the leftmost site (i.e., site 1), the first line of Eq. \eqref{aeq:recurrence} can recurrently yield the amplitude $\psi_n(\varGamma)$ for  $n\in[2,m]$. Taking $E$ as another given parameter, $\psi_n$ is also an explicit polynomial function of $E$, i.e., $\psi_n=\psi_n(E,\varGamma)$; the same for the amplitude of the last non-zero amplitude $\psi_m (E,\varGamma)$.
	
	To solve $E$ in terms of $\varGamma$, we numerically find in Fig. \ref{fig1}~(a) in the main text $E\rightarrow 0$ when $\varGamma\rightarrow 0$. Thus, the second line of Eq. \eqref{aeq:recurrence} can yield
	\begin{eqnarray}
		0&=&E-\varGamma\psi_m^2(E,\varGamma) \notag\\
		&\approx& E-\varGamma[C(\varGamma)+D(\varGamma)E]=E[1-\varGamma D(\varGamma)]-\varGamma C(\varGamma)\notag\\
		&\approx&E-\varGamma C(\varGamma),
	\end{eqnarray}
	where the first approximate equality works by expanding $\psi^2_m(E,\varGamma)$ in terms of $E\ll1$ up to the first order with $C(\varGamma)$ and $D(\varGamma)$ being the coefficients of expansion,
	and the second by further considering the condition $\varGamma\ll 1$.
	The expression means that we just need to retain in $\psi_m^2(E,\varGamma)$ the term $C(\varGamma)$ that is independent of $E$;
	in other words, it is enough to keep only the terms independent of $E$ while executing the recursion of Eq. \eqref{aeq:recurrence}, yielding
	\begin{eqnarray}
		\psi_m&=& (-\varGamma)\psi_{m-1}^3 \notag\\
		&\approx&(-\varGamma)[(-\varGamma)\psi_{m-2}^3]^3=(-\varGamma)^{1+3}\psi_{m-2}^{3^2} \notag\\
		&\approx&(-\varGamma)^{3+1}[(-\varGamma)\psi_{m-3}^3]^{3^2}=(-\varGamma)^{1+3+3^2}\psi_{m-3}^{3^3} \notag\\
		&&\cdots\notag\\
		&\approx&(-\varGamma)^{\beta_m}(\psi_1)^{3^{m-1}},
	\end{eqnarray}
	where $\beta_m=\sum_{n=0}^{m-2}3^n=(3^{m-1}-1)/2$.
	As a result, from the second line of Eq. \eqref{aeq:recurrence} we have the energy $E_m(\varGamma)$ of the SDB maximally spreading to the $m$th site [i.e., $\psi_m(\varGamma)\ne 0$] as follows:
	\begin{eqnarray}
		E_m(\varGamma)= \varGamma \psi_m^2(\varGamma)
		\approx [\varGamma\psi_1^{2}(\varGamma)]^{3^{m-1}}.
	\end{eqnarray}
	
	As for the wave function, because $E_m(\varGamma)$ has the same order of $\varGamma$ as $\varGamma\psi_m^2(\varGamma)$, which possesses the order of $\varGamma$ not less than $\varGamma\psi_{n}^2(\varGamma)~(n<m)$, we can safely discard $E_m(\varGamma)$ in each parenthesis of the first line of Eq. \eqref{aeq:recurrence}.
	Thus, we have the wave function corresponding to $E_m(\varGamma)$ is
	\begin{equation}
		\psi_n(\varGamma)\approx(-\varGamma)^{\frac{3^{n-1}-1}{2}}[\psi_1(\varGamma)]^{3^{n-1}}\,(n=1,\cdots, m),
		\label{aeq:wavefunction}
	\end{equation}
	which is consistent with $\psi_1(\varGamma)$ and $\psi_m(\varGamma)$. $\psi_1(\varGamma)$ can be further determined by the normalization condition 	\begin{eqnarray}
	1&=&\sum_{n=1}^{m}\psi_n^2(\varGamma)\approx\sum_{n=1}^m(-\varGamma)^{3^{n-1}-1}[\psi_1^2(\varGamma)]^{3^{n-1}}\notag\\
	&=&(-\varGamma)^{-1}\sum_{n=1}^m[(-\varGamma)\psi_1^2(\varGamma)]^{3^{n-1}}\notag\\
	&\approx&(-\varGamma)^{-1}[(-\varGamma)\psi_1^2(\varGamma)]=\psi_1^2(\varGamma),
	\end{eqnarray}
where we only keep the leading term of the summation for the last approximate equality due to $|-\varGamma\psi_1^2(\varGamma)|\ll 1$.
Thus, we have $\psi_1^2\approx 1$, which means that the wave function still aggregates, as expected, to the left boundary, like the unique eigenstate at the linear EP.

Finally, we have the energy and the corresponding (unnormalized) wave function at $\varGamma\ll 1$ as
\begin{equation}
	\left\{
	\begin{aligned}
	E_m(\varGamma)	&\approx \varGamma^{3^{m-1}}\\
	\psi_n(\varGamma)&\approx(-\varGamma)^{\frac{3^{n-1}-1}{2}}, ~~~(n=1,\cdots m)
\end{aligned}
\right. . 
\end{equation}
Here, we can see that the power-law scalings emerge from the linear high-order EP and that the wave function is double-exponentially decayed.

\bibliography{ref}

\begin{thebibliography}{62}
\bibitem [{CM()}]{CM}%
  \BibitemOpen
  \href@noop {} {}\bibinfo {note} {At or near the linear high-order exceptional
  point, the dynamics has its own scaling laws with the amplitude amplifying in
  a power-law fashion with respect to time, converging to the unique
  eigenvector at the exceptional point after a long time evolution. The
  relevant study is coming in our furture work.}\BibitemShut {Stop}%
\end{thebibliography}
\bibliographystyle{apsrev4-1}

\end{document}